\documentclass[12pt]{article}

\pdfoutput=1

\usepackage[american]{babel}
\usepackage{csquotes}
\usepackage{natbib}
\usepackage{csquotes,amsmath, amsthm, amssymb, amsbsy, mdwlist}

\usepackage[belowskip=-5pt,aboveskip=0pt]{caption}
\usepackage[hyperfigures=true]{hyperref}
\newcounter{blind}
\newcommand{\blind}[2]{\ifthenelse{\equal{\value{blind}}{0}}{#1}{#2}}
\hypersetup{
pdfauthor = {Mark S. Handcock;Krista J. Gile; Corinne M. Mar},
pdftitle = {Estimating Hidden Population Size using Respondent-Driven Sampling Data},
pdfkeywords = {Hard-to-reach population sampling; Network sampling; Social networks; Successive sampling; Model-based survey sampling, Probability proportional to size without replacement sampling, PPSWOR}
}

\usepackage{ifthen}
\usepackage[usenames]{color}
\newcounter{content}
\newcounter{notes}
\newcounter{techreport}

\usepackage{amssymb,amsmath,amsthm}
\usepackage{graphicx}
\usepackage{mathrsfs} % used for mathscr for loglikelihood
\usepackage{amsmath}
\usepackage{amssymb}
\usepackage[all]{xy}
\usepackage{color, graphics}
\usepackage{palatino, url, multicol}
\usepackage{graphicx}
\usepackage{subfigure}
\usepackage{mathrsfs}
\usepackage{multicol}
\usepackage{tabularx}
\newcommand{\bi}{\begin{itemize}}
\newcommand{\ei}{\end{itemize}}
\definecolor{Emphcolor}{cmyk}{0,0.89,0.94,0.1}
\definecolor{Netcolor}{rgb}{.8,0,.9}
\definecolor{Diseasecolor}{rgb}{1,.8,.2}
\definecolor{Sampcolor}{rgb}{0,.9,.3}
\definecolor{Black}{rgb}{0,0,0}
\definecolor{Red}{rgb}{1,0,0}
\definecolor{Blue}{rgb}{0,0,1}
\definecolor{Gray}{gray}{.6}

\newcommand{\IR}{\mathbb{R}}

\font\elevenrm=cmr11
\newcommand{\ql}{{\elevenrm ``}}
\newcommand{\qr}{{\elevenrm "}\ }
\newcommand{\qrns}{{\elevenrm "}\kern-2pt}

\theoremstyle{plain}

\theoremstyle{definition}

\graphicspath{{../figures/}}

\usepackage{graphicx}
\usepackage{float}
\usepackage{rotating}
\usepackage{color}
\usepackage{hypernat}
\usepackage{pdflscape}

\usepackage{graphics}
\usepackage{subfigure}
\usepackage[all]{xy}
\usepackage{amsmath}
\usepackage{amssymb}
\usepackage[all]{xy}
\usepackage{palatino, url, multicol}
\usepackage{mathrsfs}
\usepackage{multicol}
\usepackage{enumerate}

\definecolor{Emphcolor}{rgb}{.1,.1,.5}
\definecolor{Red}{rgb}{.9,0,.1}
\definecolor{Blue}{rgb}{.1,.1,.5}

\newcommand{\bea}{\begin{eqnarray}}
\newcommand{\eea}{\end{eqnarray}}
\usepackage{ifthen}
\newboolean{draft}
\setboolean{draft}{false}% to ignore the knotes
\newcommand{\knote}[1]{\ifthenelse{\boolean{draft}}{{\bf knote:~}{\it
#1}\relax}{}}
\newcommand{\mnote}[1]{\ifthenelse{\boolean{draft}}{{\bf mnote:~}{\it
#1}\relax}{}}

\title{Estimating Hidden Population Size using Respondent-Driven Sampling Data}
\author{\blind{Mark S. Handcock\footnote{Professor of Statistics, Department of
Statistics, University of California, Los Angeles, CA 90095-1554 (E-mail: \emph{handcock@ucla.edu}).} \and Krista J. Gile\footnote{Assistant Professor of Statistics,
  Department of Mathematics and Statistics,
  University of Massachusetts, Amherst,
  MA 01003-9305 (E-mail: \emph{gile@math.umass.edu}).} \and Corinne M. Mar\footnote{CSDE, University of Washington, Seattle, WA 98195-1554 (E-mail: \emph{cmmar@uw.edu}).}}{}}
\date{} % To turn the date off

\makeatletter
\renewcommand{\@maketitle}{
\newpage
 \null
 \vskip -40pt%
 \begin{center}%
 \vskip -70pt%
  \let \footnote \thanks
  {\LARGE \@title \par}%
    \vskip 1.5em%
    {\large
      \lineskip .5em%
      \begin{tabular}[t]{c}%
        \@author
     \end{tabular} 
      \vskip 1.5em%
         September 27, 2012
      \par}%
 \end{center}%
 \vskip -20pt%
 \par} \makeatother

\begin{document}
\maketitle
\thispagestyle{empty}
\setcounter{page}{0}

\begin{abstract}
Respondent-Driven Sampling (RDS) is 
an approach to sampling design and inference in hard-to-reach human populations. 
Typically, a sampling frame %for the target 
%population 
is not available, and population members are difficult to identify or recruit from broader sampling frames.
Common examples %of such populations in a behavioral and social setting 
include injecting drug users, men who have sex with men, and female 
sex workers.  Most analysis of RDS data has focused on estimating aggregate
characteristics, such as disease
prevalence.  However, RDS is often conducted in settings where
the population size is unknown and of great independent interest.
This paper presents an approach to estimating the size of a
target population based on data collected
through RDS.  

The proposed approach uses a successive sampling
approximation to RDS to
leverage information in the ordered sequence of observed personal
network sizes.
The inference uses the Bayesian framework, allowing for the incorporation
of prior knowledge. % about population size % that allows
A flexible class of priors for the population size is 
proposed that aids elicitation. 
An extensive simulation study
provides insight into the performance of the method for estimating population size under a broad range of 
conditions.
A further study shows the approach
also improves estimation of aggregate characteristics.
A particular choice of the prior produces interval
estimates with good frequentist properties. 
Finally, the method demonstrates sensible results when used to estimate the numbers of sub-populations
most at risk for HIV in two cities in El Salvador.

This article has supplementary material online.
\end{abstract}
\blind{}{
\vspace*{.3in}
}

\newcommand{\N}{\mathbb{N}}
\newcommand{\s}{\mathscr{S}}

\newcommand{\q}{Q}
\newcommand{\qdn}{Q_{\N,n}}
\newcommand{\qys}{\q^*_{k,z}(y,\s)}
\newcommand{\qysk}{\q^*_{k,z}(y^i,\s^i)}
\newcommand{\mI}{\hat{\mu}_I}
\newcommand{\hatmu}{\hat{\mu}}
\newcommand{\mush}{\hat{\mu}_{SH}}

\newcommand{\p}{{\bf p}}
\newcommand{\vp}{{\bf p}}
\newcommand{\D}{{\bf D}}
\newcommand{\X}{{\bf X}}
\newcommand{\Z}{{\bf Z}}
\newcommand{\vZ}{{\bf Z}}
\newcommand{\Y}{{\mathscr{Y}}}
\newcommand{\vd}{{\bf d}}
\newcommand{\z}{{\bf z}}
\newcommand{\vz}{{\bf z}}
\newcommand{\vs}{{\bf s}}
\newcommand{\vS}{{\bf S}}
\newcommand{\vV}{{\bf V}}
\newcommand{\vv}{{\bf v}}

\newcommand{\zs}{{\bf z_s}}
\newcommand{\ds}{{\bf d_s}}  %could also consider seed degree, but maybe not so helpful?
\newcommand{\pc}{{\bf p_c}}

\newcommand{\mhs}{\hat{\mu}_{MA}}

\abovedisplayshortskip=2pt
\belowdisplayshortskip=2pt
\abovedisplayskip=2pt
\belowdisplayskip=2pt

\advance\baselineskip by -0.3pt

\section{Introduction to Respondent-Driven Sampling}

Respondent-Driven Sampling (RDS, introduced by \cite{heck97}) is 
an approach to sampling from hard-to-reach human populations in the interest of conducting statistical inference, typically on population proportions.  In such hard-to-reach populations, a sampling frame for the target population is not available, and members are difficult to identify or recruit from broader sampling frames.
In public health, RDS is often used in studies of high-risk populations such as injecting drug users, men who have sex with men, and female sex workers; 128 of these studies are summarized in  \citet{johnston08}.  RDS has also been used in other populations such as jazz musicians  \citep{JazzMusicians2001} and unregulated workers \citep{bernhardt2006}.  

RDS is a form of link-tracing network sampling, in which subsequent sample members are selected from among the social relations of current sample members.  Unlike most link-tracing designs, {\it respondent-driven} sampling relies on study respondents to choose which of their contacts will be sampled next.  Each respondent is given a small number of uniquely identified coupons to distribute among their contacts in the target population.   Contacts receiving coupons become eligible for the study.

Most existing estimators from RDS data attempt to estimate population 
proportions\newline
\citep{heck97,Heckathorn2002,salgheck04,volzheck08,gilesppsjasa09,gilehan11MA}. Population size estimation based on RDS data is also of interest for three reasons:  First, these data are often collected in precisely the populations in which there is interest in population size.  In fact, RDS-based prevalence estimates are often used in the Estimation and Projection Package (EPP) model used by UNAIDS \citep{unaids2009EPPconc}.  For concentrated epidemics, EPP estimates national HIV rates based on both prevalence and population size estimates for several high-risk populations.  The resulting estimates of the numbers of infections are used in decisions about
resource allocation, research design and intervention planning \citep{unaids:2010size}.
Second, new prevalence estimators for RDS \citep{gilesppsjasa09, gilehan11MA} require estimates of the size of the population.  And finally, because the information in the sequence of RDS samples has not yet been exploited to estimate population size, this approach introduces a new source of information on the size of the hard-to-reach population.

There are many approaches to estimating the size of a hard-to-reach human 
population \citep{unaids:2010size,bao:raftery:reedy:2010,Pazbailey2011,BerchenkoFrost11},
including a few that use link-tracing samples similar to RDS \citep{franksnijders94, felixmedinathompson04}.  
The validity of these approaches typically rests on strong assumptions about the
populations and adherence to sampling designs.
A common approach is to use capture-recapture sampling
\citep{fienbergjohnsonjunker1999,Pazbailey2011,rocchetti2011}, which estimates population size based on the overlap between 
two or more captures of population members. Related approaches include the (network) scale-up or
multiplier methods. In these, the captures may not be based on a 
sampling design but may be from enumerations or convenience samples. The
accuracy of multiplier methods varies with the quality of the captures
data \citep{unaids:2010size,Salganik15112011}.
In particular, \citet{fienbergjohnsonjunker1999} present a general approach to
population size estimation using multiple-recapture data  and develop a
sophisticated Bayesian estimation approach for it.
Because they are based on multiple captures, to date, all methods that use RDS data require additional data collected by mechanisms other than RDS 
\citep{johnstonetal2011,Bernard01122010,Niccolai26082010,Salganik15112011}. 
The primary contribution of this paper is to introduce an estimator of population size based solely on RDS data. 
This approach is also novel in that the population size estimates are based on information in the sample sequences, exploiting the dependence in the sampling process.  In contrast, most inference from sampled data relies directly on the sampled values and treats dependence in the sample as a nuisance.
The proposed estimator can also be combined with estimates based on other
approaches to produce improved inference about the population size.

The proposed approach is founded on the successive sampling approximation to the RDS process introduced in \citet{gilesppsjasa09}.  It extends the approach developed by \citet{west96}
for ecological applications (e.g., estimating the number of oil
fields based on the sizes of the known fields).  Under successive sampling, larger units tend to be sampled earlier.  This approach therefore leverages the information in the decreasing size of sampled units (a function of oil reserve magnitude for West's application, and social connectedness for RDS) over time to make inference about population size.  Like West, it uses   a super-population model-based formulation within 
a Bayesian inferential framework by positing a prior
distribution over population size.  It differs from that of West in three key ways:
First, the unit sizes are modeled as discrete rather than continuous. Second, the branching and network nature of the RDS sample may
reduce or confound the information in the ordering of the sample.
Third, the sample sizes of RDS samples are typically larger,
and with a different range of unit sizes than in the data available in
ecological applications such as oil fields.

The next section (Section 2) develops the inferential framework and a flexible class of priors for the population size.
In Section 3, a simulation study illustrates the frequentist 
performance of the population size
estimator derived from the Bayesian framework, as well as the performance of a 
prevalence estimator based on this estimator.  
Section 4 applies the proposed method to data collected on two most-at-risk
populations in El Salvador.
Section 5 concludes the paper with a broader discussion.

\section{Bayesian Inference for the Population Size}\label{overallmodelinference}

The goal here is to make inference for population size $N$.
The approach taken is Bayesian, treating $N$ as an unknown parameter.  This requires a probability model for the observed data given $N$, as well as a prior for $N$.  Most information about the population size is drawn from the pattern in the sampling process.  In particular, 
this sampling model is non-amenable to the model \citep{hangile10aoas}.  For this reason, the probability model must represent both the sampling structure, and a superpopulation model.

The distribution of the sampling process of the units is modeled as a function of
their {unit sizes}.  % represented by $u_1, u_2, \ldots, u_N$.  
The sampling model, described in Section \ref{ssinference} below, follows \citet{gilesppsjasa09} and is based on a successive sampling approximation to the RDS process.  
The super-population model for these unit sizes is given in Section \ref{sdm}.
A likelihood function for $N$ can be computed from these two models and then combined with a prior to make inference for $N$.

The inferential frame is described as follows:  Section \ref{sizemodelinference} introduces the form of the likelihood.   Section \ref{ssinference} adds the particular form of the sampling distribution based on successive sampling. %approximation to RDS. 
 Section \ref{sec:bayesdegree} introduces the Bayesian frame for inference for the parameter of the unit size distribution, which is extended in Section \ref{sec:bayessize} to inference for $N$.  Section \ref{sdm} presents the parametric model for the unit size distribution.  %Section XX combines these pieces to present our inferential approach for population size.  
Finally, Section \ref{priors} presents the forms of the prior distributions for
the super-population model and the population size.

\subsection{Likelihood for the Super-population Parameter}\label{sizemodelinference}
\newcommand{\iid}{{\stackrel{\rm i.i.d.}{\sim}}}
Consider a population of $N$ units, 
denoted by indices $1,\ldots, N$ with an associated variable 
{\it unit size} represented by $U_1, U_2, \ldots, U_N$. 
For RDS, unit sizes are often the numbers of network connections, also known as personal network sizes or {\it degrees},  but they can be any function of individual unit variables.
The unit sizes are treated as an i.i.d. sample of size $N$
generated from a super-population model based on some (unknown) distribution. % $f.$
For simplicity of presentation, 
the unit sizes are presumed to have the natural numbers as their support 
(e.g., degrees). Specifically:
$U_i \iid f(\cdot|\eta)$
where $f(\cdot|\eta)$ is a probability mass function (PMF) with support
%$s=
$1,\ldots,$ and $\eta$ is a parameter. % (See Handcock and Jones etc).

\newcommand{\bs}{\backslash}
\newcommand{\bsg}{\backslash g}

Consider first a general ordered sampling design.
The indices of the sequentially sampled units identify the order of the sample, denoted 
by the random vector $G=(G_1, \ldots, G_n)$, where realized values %The process to deal with ties
$g=(g_1, \ldots, g_n)$ represent the observed indices of the successively
sampled units.  Let $\bsg = \{1, \ldots, N\}\bs\{g_1, \ldots, g_n\}$ represent the set of indices of the unobserved population units.
Further, consider the observed and unobserved unit sizes.  Let $U_{obs} = \{U_{g_1}, U_{g_2}, \ldots, U_{g_n}\}$, %where realized values represent the 
the random vector of observed unit sizes, with values $u_{obs}=(u_{g_1},\ldots,u_{g_n})$.  Similarly, let $U_{unobs} = \{U_{i}\}_{i \in \bsg}$ and 
$u_{unobs}= \{u_i\}_{i \in \bsg}$ represent the random and realized values of the unit sizes of the unobserved units.  
Thus the full observed data are $\{g, u_{obs}\}$.

Inference for $\eta$ should be based on all the available
observed data including the sampling sequence information. 
The likelihood is any function of $\eta$ proportional to
$p(G, U_{obs} \vert \eta )$:
\begin{eqnarray}
~~&~~&L[\eta \vert U_{obs}=u_{obs}, G=g]
\propto p(G=g, U_{obs}=u_{obs} \vert \eta)\nonumber \\
&=&\sum\limits_{u_{unobs} \in {\cal U}(u_{obs})}
{p(G=g, U=(u_{obs}, u_{unobs}) \vert \eta )}\nonumber\\
&=&\sum\limits_{u_{unobs} \in {\cal U}(u_{obs})}
{p(G=g \vert U=(u_{obs}, u_{unobs}))}{
\prod_{j=1}^{n}{f(u_{g_j} | \eta)}\cdot
\prod_{j\in \bsg}{f(u_j | \eta)}}, \label{fullikelihood}
\end{eqnarray}
where ${\cal U}(u_{obs})$ is the set of possible $u_{unobs}$ given $u_{obs}$, that is the %sequences of 
unit sizes possible for the remaining $N-n$ units given that the first $n$ sampled were $u_{obs}.$ Typically, this will be the $N-n$ product support of $f(\cdot | \eta).$
Thus the correct model is related to the complete data model through the
sampling design as well as the super-population model.

\subsection{Likelihood Under Successive Sampling}\label{ssinference}

Following  \citet{gilesppsjasa09}, the RDS sampling is approximated as a successive sampling process.  Gile argues that %for certain network structures, 
this model approximates a without-replacement random walk on the network, and demonstrates that using this model can reduce finite population biases for RDS estimates of population characteristics.   This sampling scheme is also known as {\it probability proportional to size without replacement} (PPSWOR) sampling, and is treated in the survey sampling and ecological literature \citep{andkau86,nairwang89,bnw92}.  
The {\it Successive Sampling} (SS) sampling procedure is defined as follows:
\bi
\item Sample the first unit from the full population $\{1, \ldots, N\}$ with probability proportional to unit size $u_i, i=1, \ldots, N$:
$p(G_1 = k) = {u_k}/{\sum_{j=1}^{N}u_j},~~k=1, \ldots, N.$
\item Select each subsequent unit with probability proportional to unit size {\it from among the remaining units}, such that
\begin{equation}
p(G_i = k \vert G_1 = g_1, \ldots, G_{i-1} = g_{i-1})
= \left\{ \begin{array}{cl} \frac{u_k}{\sum_{j \notin \{g_1, \ldots, g_{i-1}\}}  u_j} & k \notin \{g_1, \ldots, g_{i-1}\} \\
0 & {\rm otherwise}\label{sstransition}
\end{array} \right., i = 2,\ldots,n.
\end{equation}
\ei

The probability of the observed sequence $g$
for a given population of unit sizes is:
\begin{equation*}
  \setlength{\itemsep}{-1pt}\setlength{\topsep}{0pt}\setlength{\partopsep}{0pt}\setlength{\parskip}{0pt}
%p(G_1 \ldots G_{n} | u_{1}, \ldots, u_{N}) = \frac{N!}{(N-n)!}
p(G=g | U=(u_{obs}, u_{unobs}) ) =
\frac{N!}{(N-n)!}\prod_{k=1}^{n} \frac{u_{g_k}}{r_k}
\label{bnwsequence}
\end{equation*}
where
\begin{equation}
r_{k} = \sum_{i=1}^{N}{u_{i}} - \sum_{j=1}^{k-1}{u_{g_j}} ~~~~~~~~~~k = 1, \ldots, n,
%r_{k} = \sum_{i=k}^{N}{u_{i}} = \sum_{i=k}^{n}{u_{i}} + \sum_{i=n+1}^{N}{u_i}~~~~~~~~~~k = 1, \ldots, n
\label{rk}
\end{equation}
so that the full likelihood is:
\begin{eqnarray}
~~&~~&L[\eta \vert U_{obs}=u_{obs}, G=g]\
\propto p(G=g, U_{obs}=u_{obs} \vert \eta)\label{ssfullikelihood}\\
&=&
\frac{N!}{(N-n)!}
\sum\limits_{u_{unobs} \in {\cal U}(u_{obs})}
{\prod_{k=1}^{n} \frac{u_{g_k}}{r_k}}\cdot{
\prod_{j=1}^{n}{f(u_{g_j} | \eta)}\cdot
\prod_{j\in \bsg}{f(u_j | \eta)}}.\nonumber
\end{eqnarray}
This likelihood can be the basis of maximum likelihood estimation for $\eta.$
In general, this sum will be very difficult to compute because of
the $N-n$ embedded sums over typically infinite supports of $f(\cdot|\eta)$.

\mnote{The MLE of $\eta$ may be ok - but that of $N$ looks bad. Krista has done
a lot of work on this. We can show this work in an application where the
Bayesian solution works.}

Note that this likelihood involves models for both the sampling design and the super-population,  
necessary because 
the design is not amenable to the model (See the Supplementary Materials for details).
Intuitively, ignoring the sampling distribution would likely result in positive bias in inference about unit sizes as the larger-sized units will tend to be sampled first.  

\subsection{Bayesian Inference for the Unit Size Distribution}\label{sec:bayesdegree}

This section develops inference for the unit size distribution, conditional on known $N$.
\begin{equation*}
p(\eta | G=g, U_{obs} = u_{obs})
\propto
\pi(\eta){\cdot}
L[\eta \vert U_{obs}=u_{obs}, G=g],
\end{equation*}
where $\pi(\eta)$ is a prior for the unit size distribution parameter.  
For simplicity of notation, denote the observed data by $D=(U_{obs}=u_{obs}, G=g).$

Because of the complexity of computing the likelihood (\ref{ssfullikelihood}),
\citet{west96} suggests using the relatively simple 
\begin{equation}
p(G=g_{obs}, U_{obs}=u_{obs}, U_{unobs}=u_{unobs} \vert \eta )
=
\frac{N!}{(N-n)!}
\prod_{k=1}^{n}{\frac{u_{g_k}}{r_k}}\cdot{
\prod_{j=1}^{n}{f(u_{g_j} | \eta)}\cdot
\prod_{j \in \bsg}{f(u_j | \eta)}}\label{sslikelihood}
\end{equation}
and a two component Gibbs sampler with $p(\eta | D, U_{unobs} = u_{unobs})$
and $p(U_{unobs} = u_{unobs} | \eta, D).$
The current paper uses a variant of this approach for discrete unit size distributions. 

 From (\ref{sslikelihood}),
\begin{equation}
p(U_{unobs}=u_{unobs} \vert \eta, D )
\propto
\prod_{k=1}^{n}{\frac{1}{r_k}}\cdot{\prod_{j \in \bsg }{f(u_j | \eta)}}.\label{unobsposterior}
\end{equation}
As the $r_k$ are hard to deal with, \citet{west96} notes that,
$$
\frac{1}{r_k}=\int_{0}^{\infty}{e^{-{r_k}\psi_{k}}
d\psi_{k}},
$$
where $\psi_k$ has exponential distribution with rate parameter $r_k$.
That is,
\begin{equation}
p( \psi_{k}=\psi | \eta, U_{unobs} = u_{unobs}, D)
= {r_k} {\rm exp}( -{r_k}\psi )~~~~~\psi > 0, \label{fcpsi}
\end{equation}
He augments the data with $\Psi = (\psi_{1}, \ldots, \psi_{n})$ where the 
components are drawn (conditionally) independently so that
\begin{eqnarray*}
p(U_{unobs}=u_{unobs}, \Psi \vert \eta, D ) &=& 
p( \Psi=\psi | \eta, U_{unobs} = u_{unobs}, D)\cdot
p(U_{unobs}=u_{unobs} \vert \eta, D )\\
&\propto& 
\prod_{j=1}^{n}{e^{-{r_j}\psi_{j}}}\cdot
\prod_{j= \in \bsg}{f(u_j | \eta)}
\end{eqnarray*}
and from (\ref{rk}),
\begin{eqnarray}
p(U_{unobs} = u_{unobs} \vert \Psi, \eta, D)
&\propto&
p( \Psi=\psi | \eta, U_{unobs} = u_{unobs}, D)\cdot
p(U_{unobs}=u_{unobs} \vert \eta, D )\nonumber \\
&\propto& 
\prod_{i=1}^{n}{e^{-\psi_i\sum_{j= \in \bsg}{u_j}}}\cdot
\prod_{i=1}^{n}{e^{-\psi_i\sum_{j=i}^{n}{u_{g_j}}}}\cdot
\prod_{j= \in \bsg}{f(u_j | \eta)}\nonumber \\
&\propto&
\prod_{j \in \bsg}{{e^{-u_{j}\sum_{i=1}^{n}{\psi_i}}}f(u_j | \eta)}.
\label{fcdunobs}
\end{eqnarray}
Hence the elements of $U_{unobs}$ are 
conditionally an i.i.d. sample from the unnormalized PMF ${e^{-u\sum_{i=1}^{n}{\psi_i}}}f(u | \eta),$ and are, in fact,
conditionally independent of $D$.

The augmented posterior:
\begin{equation}
p(\eta, U_{unobs} = u_{unobs}, \Psi | D)
\label{augpost}
\end{equation}
can then be easily computed via a three component
Gibbs sampler.
Details of this and an explicit statement of the MCMC algorithm are given in the Supplementary
Materials.

The algorithm produces samples from
$p(\eta \vert D)$ and 
from the posterior predictive distribution for the unobserved unit sizes:
$p(U_{unobs} = u_{unobs} \vert D)$.
These in turn enable inference for such quantities as the mean unit size, the
unit size distribution, etc.

\subsection{Estimating the Size of the Hidden Population}\label{sec:bayessize}

When $N$ is unknown, it becomes an additional parameter to be estimated. For
simplicity, specify that $N$ and $\eta$ are {\sl a priori} independent so
that $\pi(N, \eta) = \pi(N){\cdot}\pi(\eta).$
A variant of the approach in the last section allows draws from
the joint posterior 
\begin{equation}
p(N, \eta, U_{unobs} = u_{unobs} | D).
\label{expostN}
\end{equation}
This change requires
$p(N, \eta, \Psi | D)$.
Using (\ref{sslikelihood}) and (\ref{fcpsi}),
\begin{eqnarray}
&~~&p(D, U_{unobs} = u_{unobs} \vert N, \Psi, \eta)\\
&\propto&
p( \Psi=\psi | N, \eta, D, U_{unobs} = u_{unobs})\cdot
p(D, U_{unobs}=u_{unobs} \vert N, \eta )\nonumber \\
&\propto& 
\frac{N!}{(N-n)!}\cdot
\prod_{i=1}^{n}{u_{g_j}{f(u_{g_i} | \eta)}}
\prod_{i=1}^{n}{e^{-\psi_i\sum_{j= \in \bsg}{u_j}}}\cdot
\prod_{i=1}^{n}{e^{-\psi_i\sum_{j=i}^{n}{u_{g_j}}}}\cdot
\prod_{j= \in \bsg}{f(u_j | \eta)}\nonumber \\
&\propto&
\frac{N!}{(N-n)!}\cdot
\prod_{i=1}^{n}{u_{g_j}{f(u_{g_i} | \eta)}e^{-\psi_i\sum_{j=i}^{n}{u_{g_j}}}}
\prod_{j \in \bsg}{{e^{-u_{j}\sum_{i=1}^{n}{\psi_i}}}f(u_j | \eta)}.
\label{fgd}
\end{eqnarray}

The full-conditional for $N$ is
\begin{eqnarray}
~~&~~&p(N \vert \eta, \Psi, D)
\propto \pi(N)p(D \vert N, \eta, \Psi)
= \pi(N) \sum\limits_{u_{unobs} \in {\cal U}(u_{obs})}
{p(D, U_{unobs} = u_{unobs} \vert N, \Psi, \eta)}\nonumber \\
&\propto& \frac{N!}{(N-n)!}\cdot\pi(N) \sum_{u_{unobs} \in U(u_{obs})}
\left\{
\prod_{j \in \bsg}{{e^{-u_{j}\sum_{i=1}^{n}{\psi_i}}}f(u_j | \eta)}
\right\}\nonumber \\
&\propto& \frac{N!}{(N-n)!}\cdot\pi(N) \prod_{j=n+1}^{N}
\left\{\sum_{v_{j}=1}^{\infty}{
e^{-v_{j}\sum_{i=1}^{n}{\psi_i}}}f(v_j | \eta)
\right\} \nonumber \\
&\propto& \frac{N!}{(N-m)!}\cdot\pi(N)\left[\gamma(\sum_{i=1}^{n}{\psi_i}, \eta)\right]^{N-n},~~~~~~~~{\rm where}~~~~\gamma(\alpha, \eta) = \sum_{j=1}^{\infty}{
e^{-{\alpha}j}f(j | \eta)}.\label{fcm}
\end{eqnarray}
The other full-conditionals are unchanged.
This leads to a four component Gibbs sampler,
the details of which are given in the Supplementary
Materials.
The algorithm can be run to produce a large sample from the augmented posterior:
$p(N, \eta, U_{unobs} = u_{unobs}, \Psi | D)$.
This can then be marginalized to produce samples from
$p(N \vert D),$
$p(\eta \vert D),$
and the posterior predictive distribution of the unobserved unit sizes, 
$
p(U_{unobs} = u_{unobs} \vert D)
$. Hence it produces posterior predictive distributions of the full population
of unit sizes $(u_i, i=1,\ldots, N)$.
These posteriors enable inference for such quantities as the population size,
the mean unit size, the
unit size distribution, etc.

\subsection{Models for the Unit Size Distribution}\label{sdm}

This approach is general with respect to the parametric distribution of unit sizes, and 
several parameterizations are included in the 
software \blind{\citep{size}}{\citep{sizeblind}}.
The question of models for the degree distributions of social networks has been
extensively studied.
Discussion of the many alternative distributional families is given in the Supplemental Materials.
Throughout the study of the synthetic populations and application to real data
the Conway-Maxwell-Poisson distribution for the unit sizes is used. It allows both under-dispersion and
over dispersion with a single additional parameter over a Poisson
\citep{shmuelietal2005}.

\subsection{Prior specification}\label{priors}

\subsubsection{Prior for the unit size distribution model}

Each two-parameter unit size distribution, including the Conway-Maxwell-Poisson, 
can be 
parameterized in terms of its mean and
standard deviation. The prior %is the joint conjugate family: the prior 
for the mean given the standard deviation is normal and
the variance is scaled inverse Chi-squared:
\begin{equation*}
\mu | \sigma \sim N(\mu_0,\sigma/df_{\rm mean})
~~~~~~~~~~~~\sigma \sim {\rm Inv}\chi(\sigma_0; df_{\rm sigma}).
\end{equation*}
The default prior on these parameters is close to
uninformative, equivalent sample size of $df_{\rm mean}=1$ for the mean of
the unit size distribution and $df_{\rm sigma}=5$ for
the variance of the unit size distribution. 

\subsubsection{Prior for the population size}

The model allows for an arbitrary prior distribution over the 
population size ($N$).  However, this is an opportunity to choose priors that
aid elicitation of expert prior information or easily incorporate previous or
concomitant sources of information about the population size.

\begin{figure}[h]
\begin{center}
    \includegraphics[width=2.5in,height=2.5in]{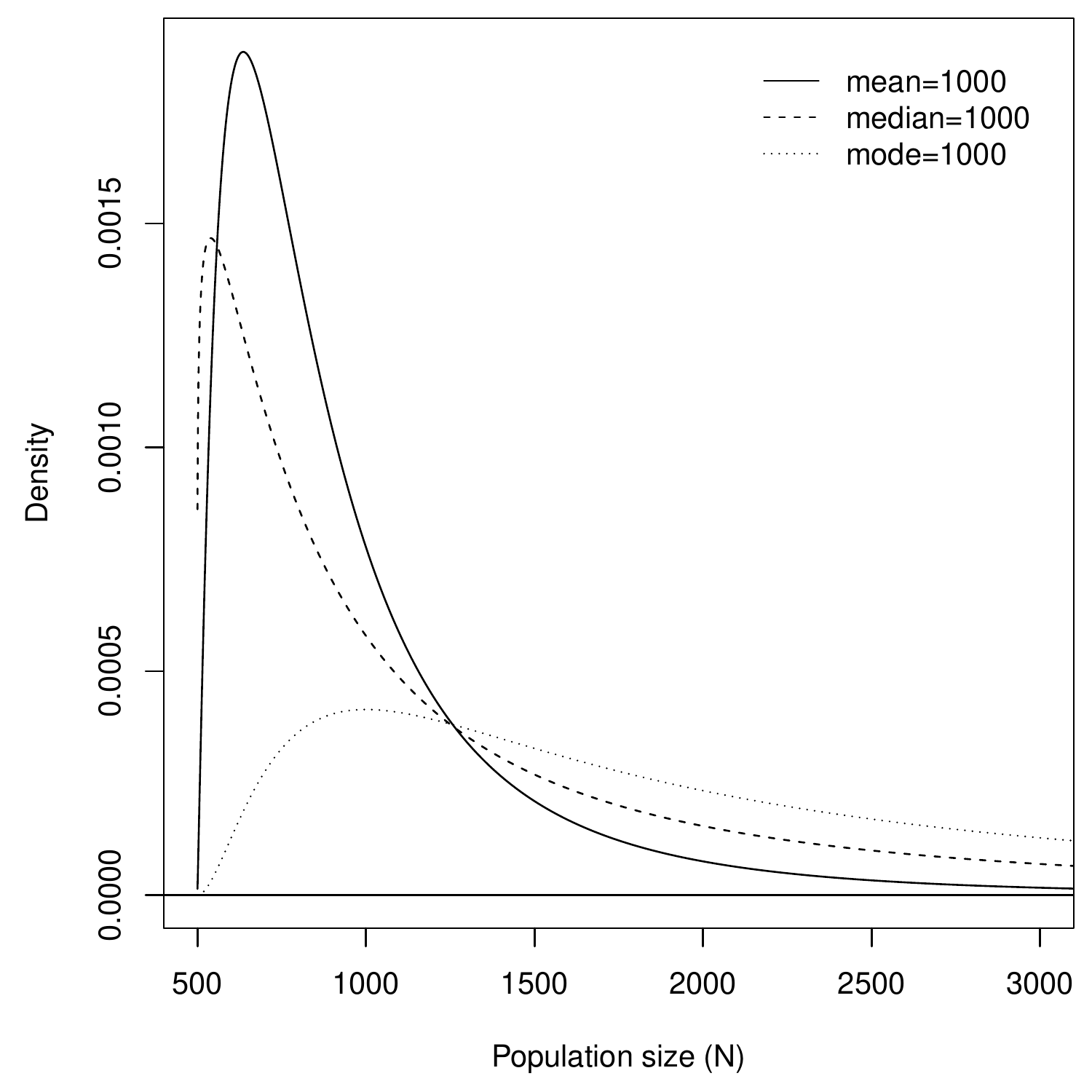}
\end{center}
\caption{Three example prior distributions for the population size 
($N$). They correspond to $\alpha=1$ and $\beta=1.55, 1.16$ and
$3$.}\label{priorplot}
\end{figure}

The most common parametric models for $N$ (e.g., Negative Binomial) typically 
have too thin tails for large $N$.
This issue has been thought through by \citet{fienbergjohnsonjunker1999}. They
suggest the prior:
\begin{equation}
\pi(N ) = (N-l)! / N! ~~~~{\rm for}~~~ n < N < N_{max},
\label{fNprior}
\end{equation}
where $N_{max}$ covers the range where the likelihood is non-negligible.
For their applications they choose their Jeffrey's prior $l=1,$
$\pi(N) \propto 1/N, n < N < N_{max}.$
In addition to these possibilities, we propose a new class of priors specifying 
knowledge about the sample proportion (i.e. $n/N$)
as a Beta$(\alpha,\beta)$ distribution.
The implied density function on $N$ (considered as a continuous variable) is:
\begin{equation}
\pi(N) = \beta{n}(N-n)^{\beta-1} / N^{\alpha+\beta} ~~~~{\rm for}~~~ N > n.
\label{nNprior}
\end{equation}
The distribution has tail behavior $O(1/N^{\alpha+1})$.
Figure \ref{priorplot} presents three different versions
of this prior, corresponding to a prior mean, median and mode of 1000.
We have found this class of priors to be very useful: 
It is often relatively flat in regions where the likelihood is centered.
The long-right-tail
allows large population sizes but the rate of decline ameliorates this. 

When $\alpha=l-1 > 0$, this class is similar to that of \citet{fienbergjohnsonjunker1999}.  The Beta prior class, however, is directly motivated as
 a proper prior on the sample proportion. 
Additional details on this prior are given in the Supplemental Materials.

\newcommand{\E}{\mathbb{E}}
\newcommand{\hatqdn}{{\hat{Q}}}
\def\E{\mathop{\rm E\,}\nolimits}

\section{A 
Simulation Study to Assess Frequentist Properties} \label{mhsim}

The primary focus of the simulation study is to evaluate the frequentist
performance of the proposed estimator for population size, including point estimation, interval estimation, and sensitivity to prior specification.  Of secondary interest is the use of these estimates to inform the estimation of population proportions using the estimator introduced in \citet{gilesppsjasa09}.

The parameters of the study are largely chosen for consistency with the simulation studies of RDS-based estimators of population proportions in
\citet{gilehansocmeth10}, \citet{gilesppsjasa09}, and \citet{tomasgile}, although with a greater range of population sizes.  
As in these studies, to increase the realism of the study,
parameters match the characteristics of the
pilot data from the CDC surveillance program \citep{aqcdc06} whenever possible.  The general procedure is: (1) 200 Networks are simulated under each test condition; (2) An RDS sample is simulated from each sampled network; (3) Point and interval estimators of population size are computed from each sample.

For the simulation, all samples are of size 500, and the population mean degree is fixed at 7. 
A discoverable class, referred to as {\ql}infection status,\qr is assigned 
to each member of the population such that each population has prevalence 20\%.

The varying characteristics of the synthetic populations are also chosen 
to represent those expected in the real world.
These characteristics include population size (i.e., the number of nodes), 
tendency for individuals to preferentially form
relations with others of the same infection status (known as {\it homophily}),
and different rates of network connectivity by infection status (referred to as
{\it differential activity}).

Population network structures are modeled using 
Exponential family Random Graph Models
(ERGM) \citep{sprh06}. That is, the $N \times N$ binary matrix of relations, $y$, is represented as a realization
of the random variable $Y$ with distribution:
\begin{eqnarray}
P_{\eta}(Y=y | x) = \exp\{\eta{\cdot}g(y,x)-\kappa(\eta,x)\}\quad \quad y\in {\cal
Y},
\label{ergm}
\end{eqnarray}
where $x$ are covariates, $g(y,x)$ is a $p$-vector of network statistics,
$\eta\in \IR^p$
is the parameter vector, ${\cal Y}$ is the
set of all possible undirected graphs, and
$\exp\{\kappa(\eta,x)\} =
\sum_{u\in{\cal Y}}\exp\{\eta{\cdot}g(u,x)\}
$
is the normalizing constant \citep{bar78}.

This modeling framework can represent a very wide range of populations, with
the particular structures determined by the choice of $g(y,x)$.
Here, differential activity is parameterized as the ratio, $\omega,$ of the mean degree of infected nodes to the
mean degree of uninfected nodes, 
where $\omega=1$ represents the absence of differential
activity.  
While homophily can and will occur on multiple variables, 
the most impactful type is that on infection status. 
Homophily is parameterized, for fixed mean degree of each group,
as the ratio, $\alpha,$ of the expected number of
discordant-infection-status ties absent homophily to the expected number of
discordant-infection-status ties with the homophily:
\begin{equation*}
 \alpha = \frac{\E(\textrm{number of infected-uninfected ties absent homophily})}{\E(\textrm{number of infected-uninfected ties})},
\end{equation*}
so that larger values of $\alpha$ indicate more homophily.
This measure is meaningful across different levels of differential activity.
Note that this parameterization of homophily is different from that in earlier studies \citep[e.g.][]{gilehansocmeth10}. 
 
These features are represented in the ERGM by choosing
network statistics to represent the mean degree, the relative
activity levels of the two groups, and homophily.   
The binary nodal covariate $x_i$ represents infection status, such that $x_i=1$ indicates infection.  These three parameters then map to the expected cell counts of the mixing matrix on infection status.  Our networks are thus simulated from an ERGM with 
\begin{equation*}
g(x,y)= \{\sum_{i=1}^N \sum_{j\neq i}  x_i x_j y_{ij}, \sum_{i=1}^N \sum_{j\neq i}  x_i (1-x_j) y_{ij}, \sum_{i=1}^N \sum_{j\neq i}  (1-x_i) (1-x_j) y_{ij}  \}.
\end{equation*}
The range of population characteristics modeled is 
(a) population size: $N \in \{5000, 1500,$ $1000, 750, 555\}$;
(b) differential activity: $\omega \in \{0.5, 1, 2\}$;
and (c) homophily: $\alpha \in \{1, 1.8 \}$.
The $\eta$ parameter of the ERGM is chosen
so the expected values of the statistics are equal to the values given
above, and the simulated networks are generated from the resulting model, as in \cite{vanduijngilehan09}.
This was implemented using the R package {\tt statnet} \citep{statnet}.

Ten seed nodes are chosen for each sample, selected (sequentially) with probability proportional to degree.  Subsequent sample waves are selected without-replacement by sampling two nodes (where possible) at random from among the unsampled alters of each sampled node.  This typically resulted in four complete waves and part of a fifth wave, stopping at sample size 500. 

Throughout the simulations, unit sizes are modeled with 
a Conway-Maxwell-Poisson distribution with diffuse priors 
(hyper parameters $\mu_0=7, df_{\rm mean}=1, \sigma_0=3, df_{\rm sigma}=5$).
The prior for the population size is a Beta distribution with  $\alpha=1$ as describe in (\ref{nNprior}).
Sensitivity of the method to incorrect priors is tested with priors based on three different prior means:
equal to the true population size, $N$, a high estimate equal to $2N$, and a low estimate halfway between $N$ and the sample size $n$:  $(N+n)/2$.    

Results from each simulation are summarized using the posterior mean as a point estimate, and the 95\% highest posterior density region as an interval estimate.

\subsection{Point and Interval Estimation of Population Size}

\begin{figure}[h]
%\begin{center}
\centering
   \includegraphics[width=5.5in]{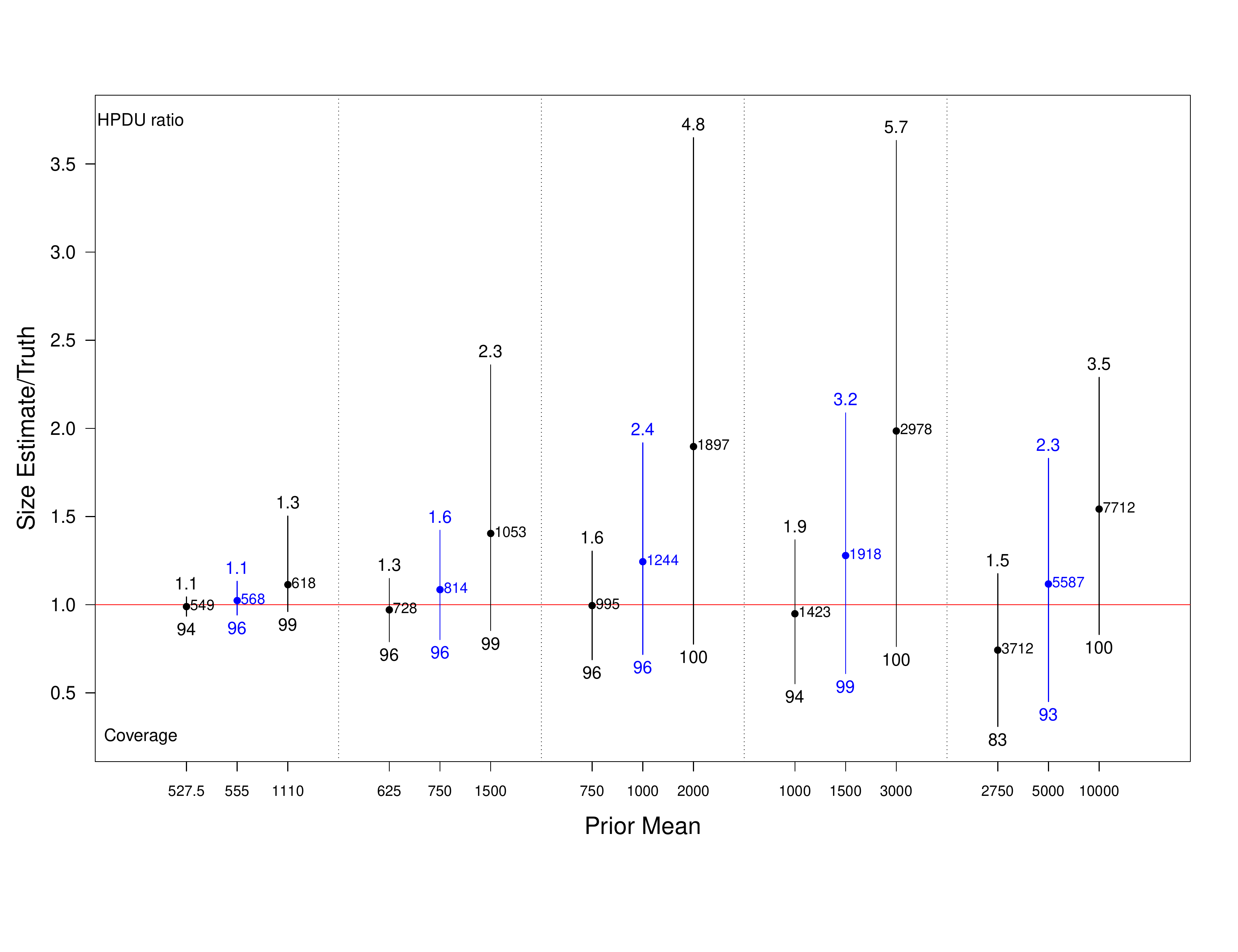}
%\end{center}
\caption{Spread of central 95\% of simulated population size estimates (posterior means) for 5 population sizes for low, accurate, and high priors.  Dots represent means.  Estimates are represented as multiples of the true population size (red line at 1 indicates true population size).  Numbers below the bars are coverage rates of 95\% HPD intervals, numbers above the bars indicate the widths of these intervals (ratio of median upper bound to truth).}\label{fig:table1}
\end{figure}

Figure \ref{fig:table1} summarizes population size estimates based on simple network structures with no homophily ($\alpha=1$) or differential activity ($\omega=1$) for all five population sizes (corresponding to different sample fractions), and low, accurate, and high prior estimates of population size.  

When the prior is correct (blue lines), average point estimates are reasonably close in all cases.  There is a small amount of positive bias.  This is because of the successive sampling (SS) approximation to the true link-tracing network sampling process.  In SS, the next unit sampled would be chosen with probability proportional to unit size from among all unsampled units.  In RDS, the network structure constrains the selection of each subsequent unit, with the effect that the decrease in sampled unit sizes over time is less sharp than in successive sampling, leading to a slight positive bias in the population size estimates.
The coverage rates of nominal 95\% credible intervals are about right in the case of accurate prior information.

Because there is limited information regarding population size in the RDS samples, results are affected by the choice of prior mean, with greater impact for smaller sample fractions.  This is because smaller sample fractions entail less exhaustion of the target population and therefore less information in the data about population size.

The coverage rates for the 95\% HPD regions for cases of prior mis-specification range from 83\% to 100\%, with higher coverage rates for higher prior means.   
These intervals can be quite wide.  Because of the lower bound induced by the sample size, interval width is largely determined by the upper limits.  The numbers above the bars in Figure \ref{fig:table1} represent the
median across each set of 200 simulations of the upper limit of the HPD intervals, represented as a multiple of the true population size. When the population size is close to the sample size, intervals are quite tight, while smaller sample fractions yield intervals that are often very large,  
with median upper point 3.2 times the true population size for $N=1500$, and 
even wider (5.7 times) with higher prior mean of $3000$. 

\subsection{Impact of Network Structure}

\begin{figure}
\centering
   \includegraphics[width=5.5in]{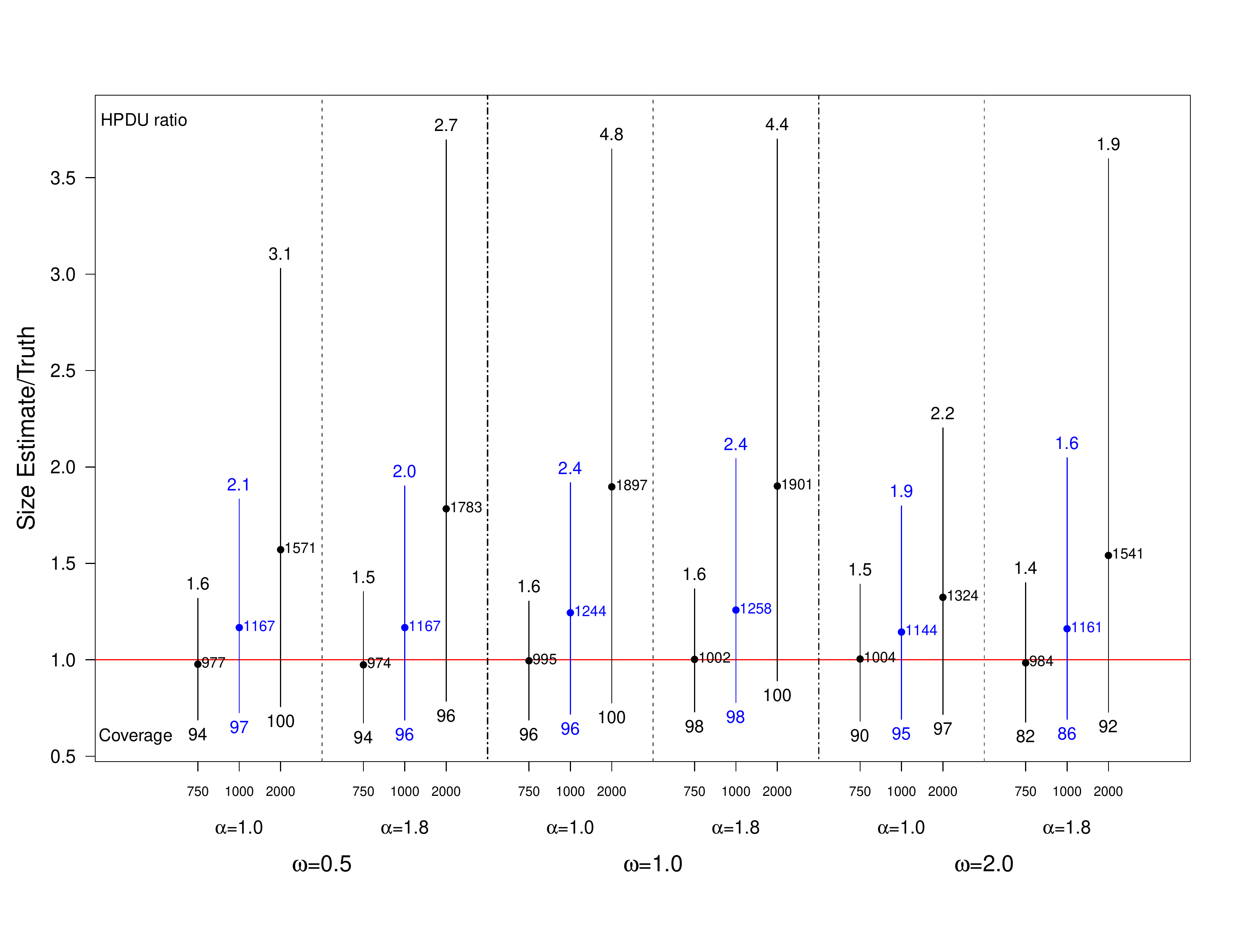}
\caption{Spread of central 95\% of simulated population size estimates (posterior means) for population size 1000 for low, accurate, and high priors, with varying levels of homophily ($\alpha$) and differential activity ($\omega$).  
The legend is the same as Figure 2.}\label{fig:table2}
\end{figure}

Figure \ref{fig:table2} summarizes the results of varying the levels of homophily and differential activity, for varying prior specifications, all for populations of size 1000.  Each pair of columns provides comparison across two levels of homophily ($\alpha$).  The paired columns are very similar, in both point and interval estimates, and across all levels of differential activity.  This suggests that under a broad range of circumstances, homophily does not have a strong first-order impact.

There does appear, however, to be a first order impact of differential activity ($\omega=0.5$ and $\omega=2$), across levels of homophily.  By systematically varying mean degree across infection groups, differential activity increases the variation in the unit size distribution, increasing the rate of decline in sampled unit sizes, and therefore providing more information about population size, resulting in better point estimates.   
Note how the prior mean 2000 cases have point estimates far closer to the truth when $\omega=2$ as compared to $\omega=1$.  The credible intervals (HPDU ratios) are also typically smaller for $\omega\not=1$ cases, without substantial reduction in the coverage rates.

\subsection{Estimation of Population Proportions}

RDS is typically conducted in the interest of estimating population features such as population proportions.  Earlier estimators based on RDS data assumed the population size was very large with respect to the sample size, so that finite population effects could be ignored.  The more recent estimator which introduces the successive sampling (SS) approximation on which this paper is based \citep{gilesppsjasa09}, however, includes a finite population adjustment, but assumes that the population size is known.  It is natural, therefore, to use the approach to population size estimation introduced in this paper to provide a population size estimate for use in the prevalence estimator in \citet{gilesppsjasa09}.  Hence this section considers estimates of infection prevalence using the SS estimator.  This section compares results using the prior mean as the population size to results using the posterior mean.  

\begin{figure}[h]
\centering
%%\begin{center}
   \includegraphics[width=5.5in]{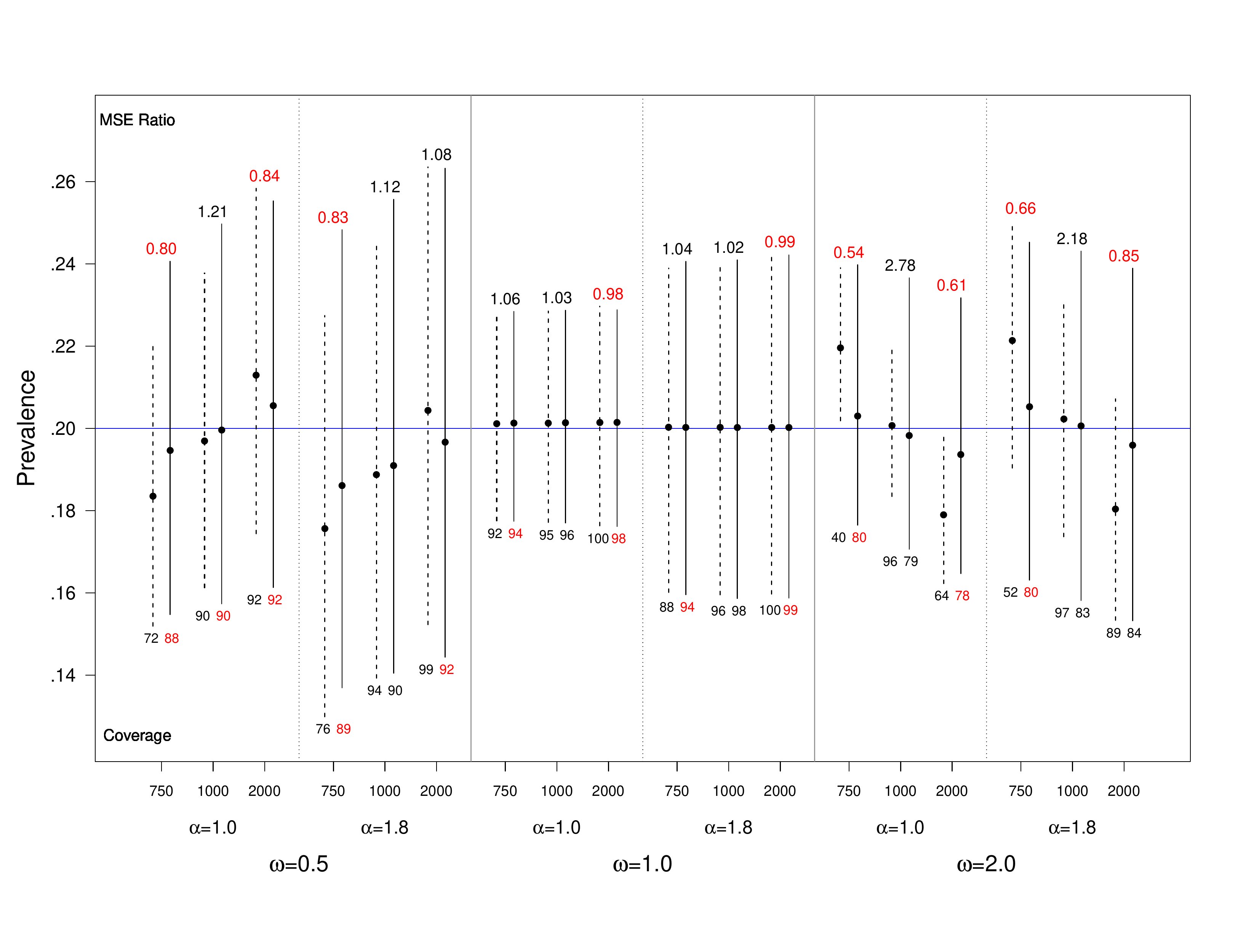}
%%\end{center}
\caption{Spread of central 95\% of simulated prevalence estimates for population size 1000, with varying levels of homophily ($\alpha$) and differential activity ($\omega$). Solid lines represent prevalence estimates based on the posterior mean, dashed lines represent comparable estimates using the prior mean.  Relative efficiency (MSE posterior/MSE prior) is given above each bar, and the coverage of nominal 95\% confidence intervals is below each bar.  The true prevalence is 0.2 (blue line).}\label{fig:table2ss}
\end{figure}

Figure \ref{fig:table2ss} shows the SS results for the same simulation conditions as Figure \ref{fig:table2}.  Absent differential activity (Figure \ref{fig:table2ss}, middle 2 columns), there is little difference between results using the prior and posterior means.  This is because the SS estimator re-weights the sample based on unit sizes (degrees) and the assumed population size.  Absent differential activity, the infected and uninfected subsamples will have similar degree distributions, and therefore be similarly affected by any aberrations in the unit weights.

In the presence of differential activity, inaccurate population size estimates introduce bias in the point estimates given by the successive sampling estimator \citep[see][]{gilesppsjasa09}.  The first and last two columns of Figure \ref{fig:table2ss} compare prevalence estimates based on the prior and posterior means for cases with strong differential activity.  Consistent with \citet{gilesppsjasa09},  the dashed bars, corresponding to estimates using the prior mean, show substantial bias when the population size is inaccurate.  This is due to imperfect adjustment for finite population effects.
The primary advantage of estimates based on the posterior mean is the reduction of this dramatic bias.  The cost of this reduction, however, is increased variance of the estimator, resulting in higher MSE in cases with small finite population biases, such as when the prior mean is correct. Note that because the bootstrap standard error estimator associated with the successive sampling estimator does not account for uncertainty in the population size, coverage rates can be dramatically low, especially for the estimator based on the prior mean.

\section{Application:  Estimating the numbers of those most at risk for HIV
in Cities in El Salvador}

Overall, El Salvador is a country with low HIV prevalence. 
As of 2010, the adult HIV prevalence was estimated at 0.8\% \citep{unaidsElsalvador2010}.  
However
the virus remains a significant threat in groups who practice 
high-risk behaviors, such as
female sex workers (FSW) and men who have sex with men (MSM)
\citep{morales2007,sotoetal2007}.

This section analyses two studies from
data collected in 2010 as part of a series of
RDS studies of populations most at risk for HIV across major El Salvadorian cities \citep{Pazbailey2011}.  

The first case is a study of FSW in the 
department of Sonsonate, which had a population of about 540,000 in
2008 \citep{guardado2010}.
This study began with 5 seeds and continued until wave 9, with
11 samples from wave 8 and 5 from wave 9, and a total of 184 samples.
The average wave number was 3.8.
A graph of the recruitment is given in the Supplementary Materials.

As is typical in these settings, the number of female sex workers in 
Sonsonate is unknown. The public health officials use
the UNAIDS national HIV estimation work group recommendation to
estimate the number of FSW based on a percent from the total adult 
female population \citep{UNAIDS2003}. 
In 2009, this group estimated that 
FSW constitute 
0.4\%-0.8\% of the urban female population 15-49 years of age (139,804)
\citep{Censos2007, Pazbailey2011}.
The range for Sonsonate is then 560 to 1120 FSW.

\begin{figure}
\centering
%\begin{center}
\subfigure
%, $Y \sim \beta$}] % caption for subfigure a
{
    \label{sonflatN}
    \includegraphics[width=2.1in]{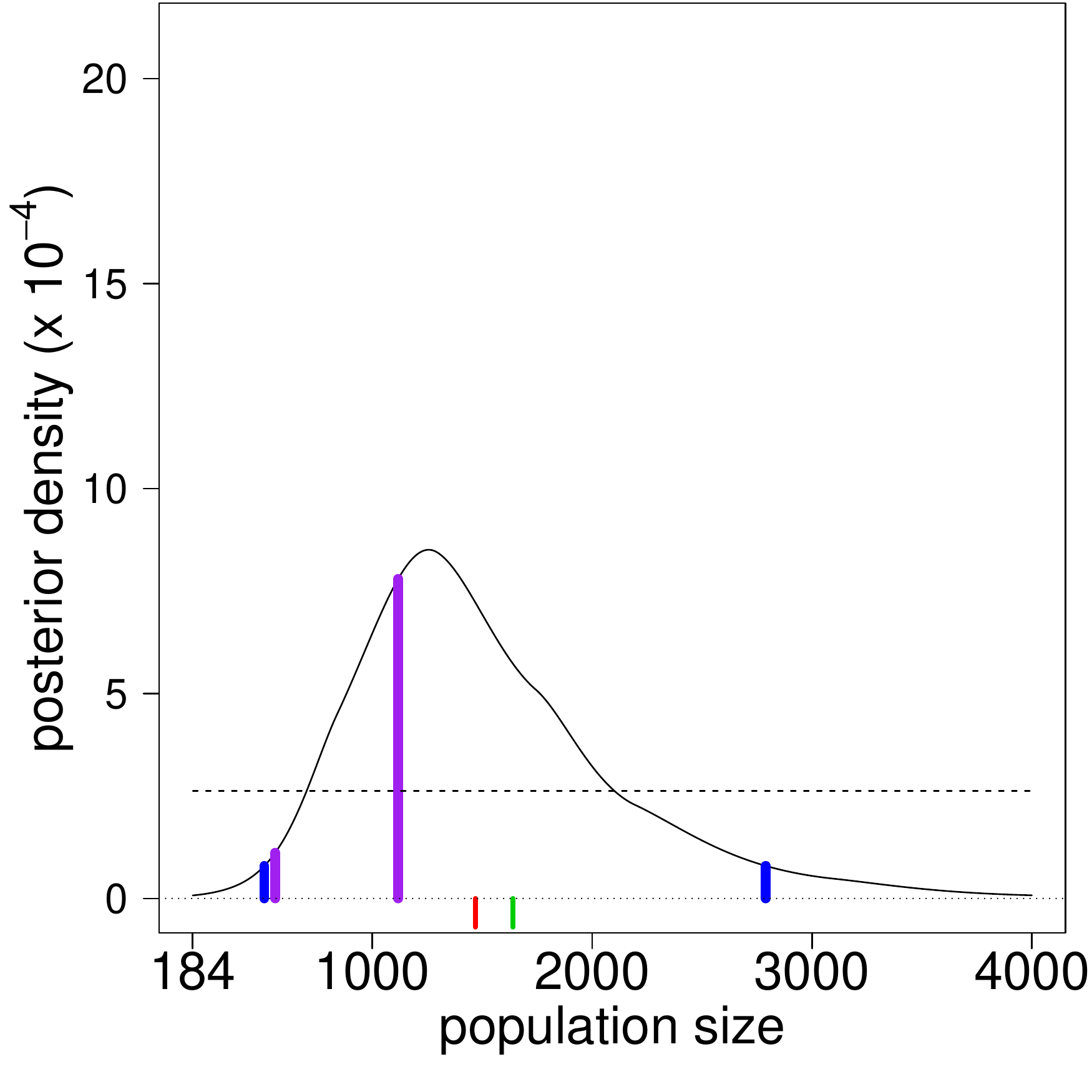}
} \hspace{-.3cm}
\subfigure
%, $Z|Y \sim \delta$}] % caption for subfigure a
{
    \label{sonmidpointN}
    \includegraphics[width=2.1in]{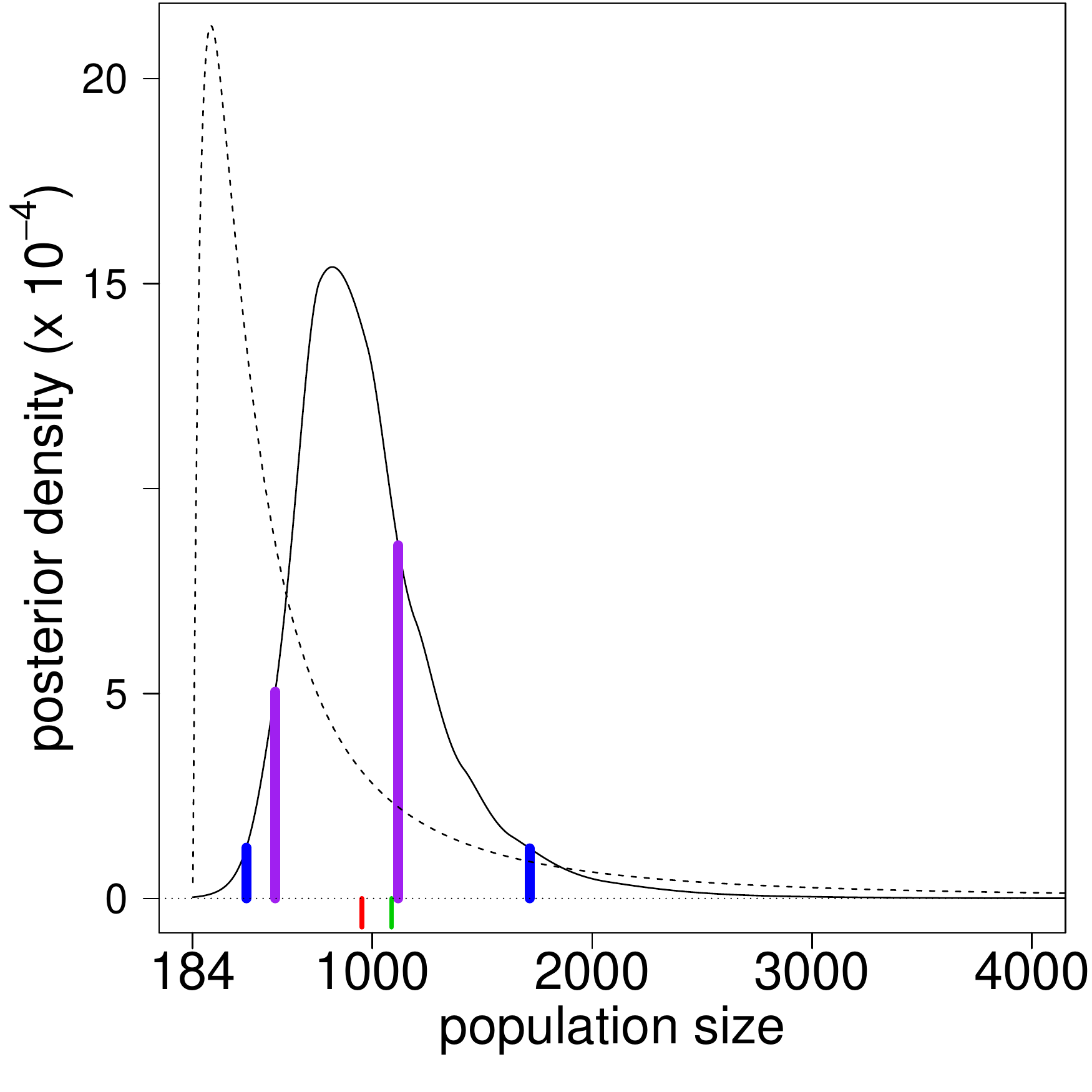}
} \hspace{-.3cm}
\subfigure
%, $Z|Y \sim \delta$}] % caption for subfigure a
{
    \label{sonUNAIDSN}
    \includegraphics[width=2.1in]{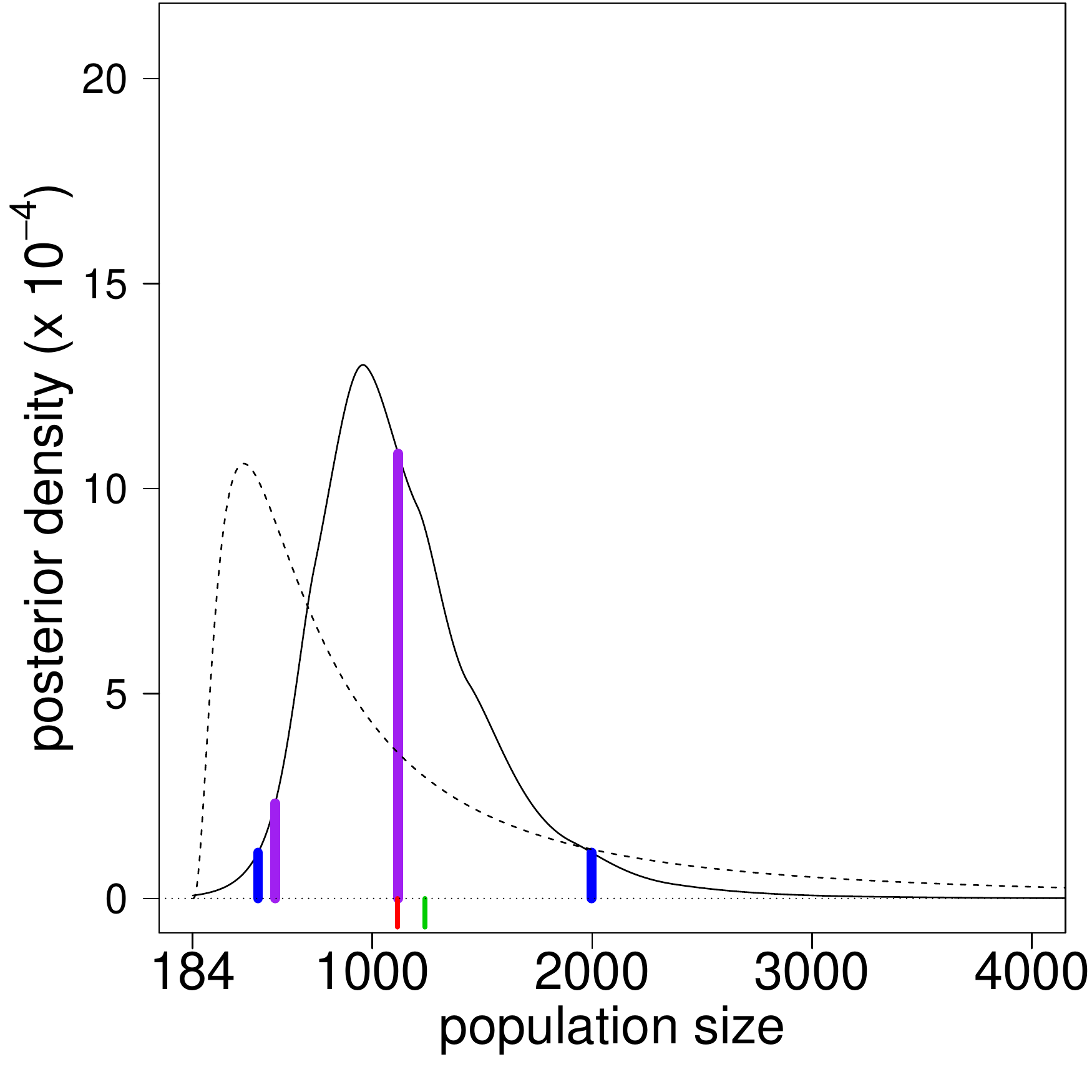}
   } 
%\end{center}
\caption{Posterior distribution for the number of female sex workers in
Sonsonate based on three prior distributions for the population size:  flat, matching the midpoint UNAIDS estimate, and interval-matching the UNAIDS estimate.
The prior is dashed.
The red mark is at the posterior median. The green mark is at 
the posterior mean. The blue lines are
at the lower and upper bounds of the 95\% highest-probability-density interval.
The purple lines demark the lower and upper UNAIDS guidelines.}
\end{figure}

Estimates of the total number of FSW in Sonsonate are calculated with three different priors.  The first prior 
is constant over the range of population sizes
where the likelihood is non-negligible.
Figure \ref{sonflatN} plots both the prior and posterior distributions.
The peakedness of the posterior shape indicates that there is information in
the data about the population size, with a mode of around 1250 FSW. The 
UNAIDS guidelines (purple lines) fall in the mid to low part of the posterior distribution, are broadly consistent with it,
and fall within the 95\% HPD interval (blue lines).

The second prior %is the reference prior used in the simulation study with 
places the prior 
mean at the mid-point of the UNAIDS suggested range
$0.6\%\times{139804}=838$.  Figure \ref{sonmidpointN} plots this prior and the resulting posterior. The mean, median and mode of the posterior fall within the UNAIDS guidelines (purple lines) and these guidelines fall within the the 95\% HPD interval.
As expected, this prior results in more mass in the posterior in the area of the UNAIDS estimates.%The 95\% interval covers the guidelines.

As UNAIDS provides a range of values, it may be useful to specify a prior based on multiple points in that range. The class of priors described by (\ref{nNprior}) allows the flexibility to choose a prior that reflects
a range of values.  In this case, two parameters $(\alpha, \beta)$ are chosen so that the prior mean is the midpoint of the range
($0.6\%$) and the lower quartile of the prior is the lower UNAIDS estimate ($0.4\%$).
Figure \ref{sonUNAIDSN} plots this prior and resulting posterior distribution.
This distribution has a mode at about 1000 FSW, and a 95\% HPD interval from
481 to 1998 FSW.

The above analysis is now repeated for a study of MSM in
San Miguel.
In 2009, the UNAIDS national HIV estimation work group estimated the number of MSM in El Salvador at 
2\%-5\% of the urban male population 15-49 years of age (148,489)
\citep{UNAIDS2003, Censos2007, Pazbailey2011}.
The range for San Miguel is then 2970 to 7425 MSM.

\begin{figure}
\centering
%\begin{center}
\subfigure
%, $Y \sim \beta$}] % caption for subfigure a
{
    \label{smflatN}
    \includegraphics[width=2.1in]{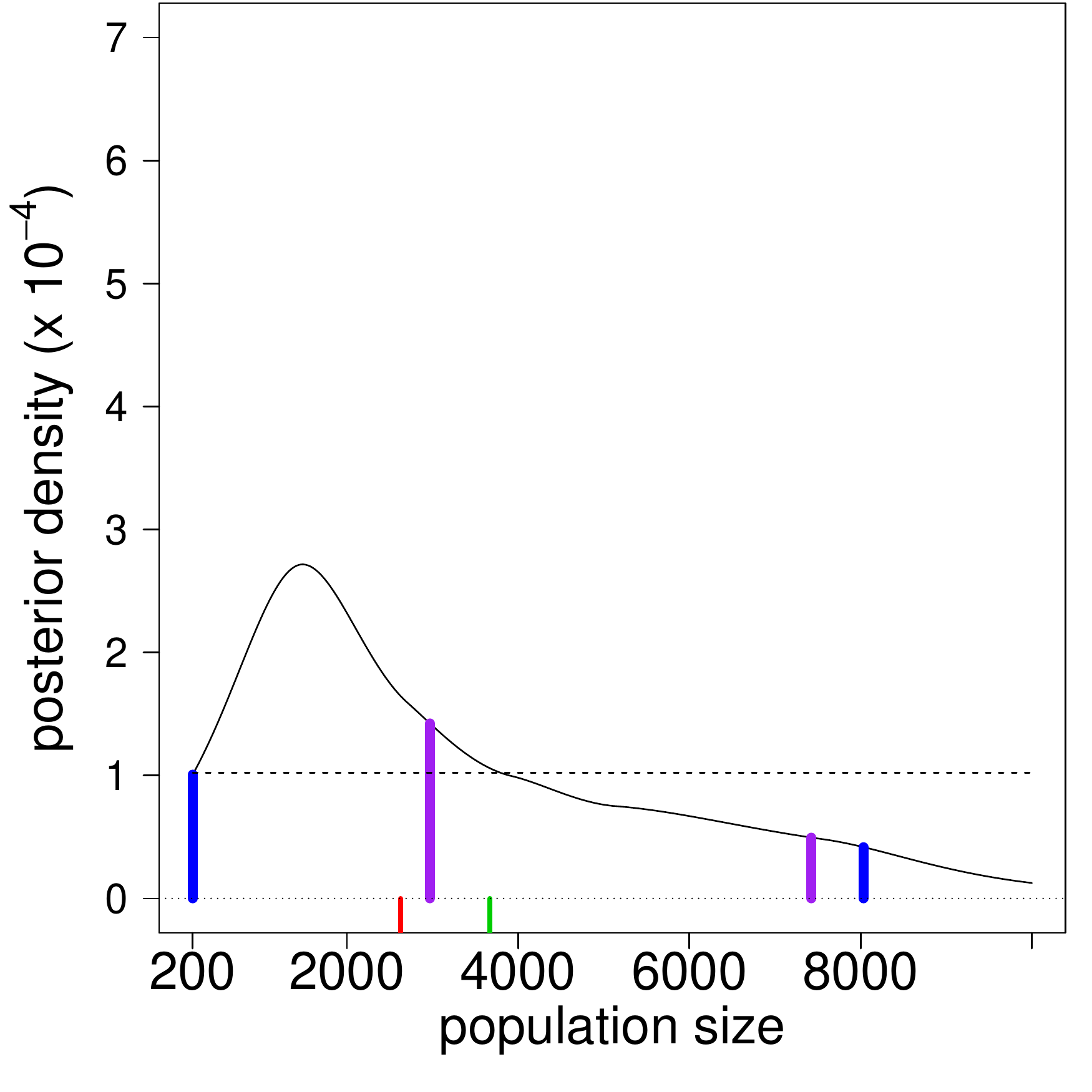}
} \hspace{-.3cm}
\subfigure
%, $Z|Y \sim \delta$}] % caption for subfigure a
{
    \label{smmidpointN}
    \includegraphics[width=2.1in]{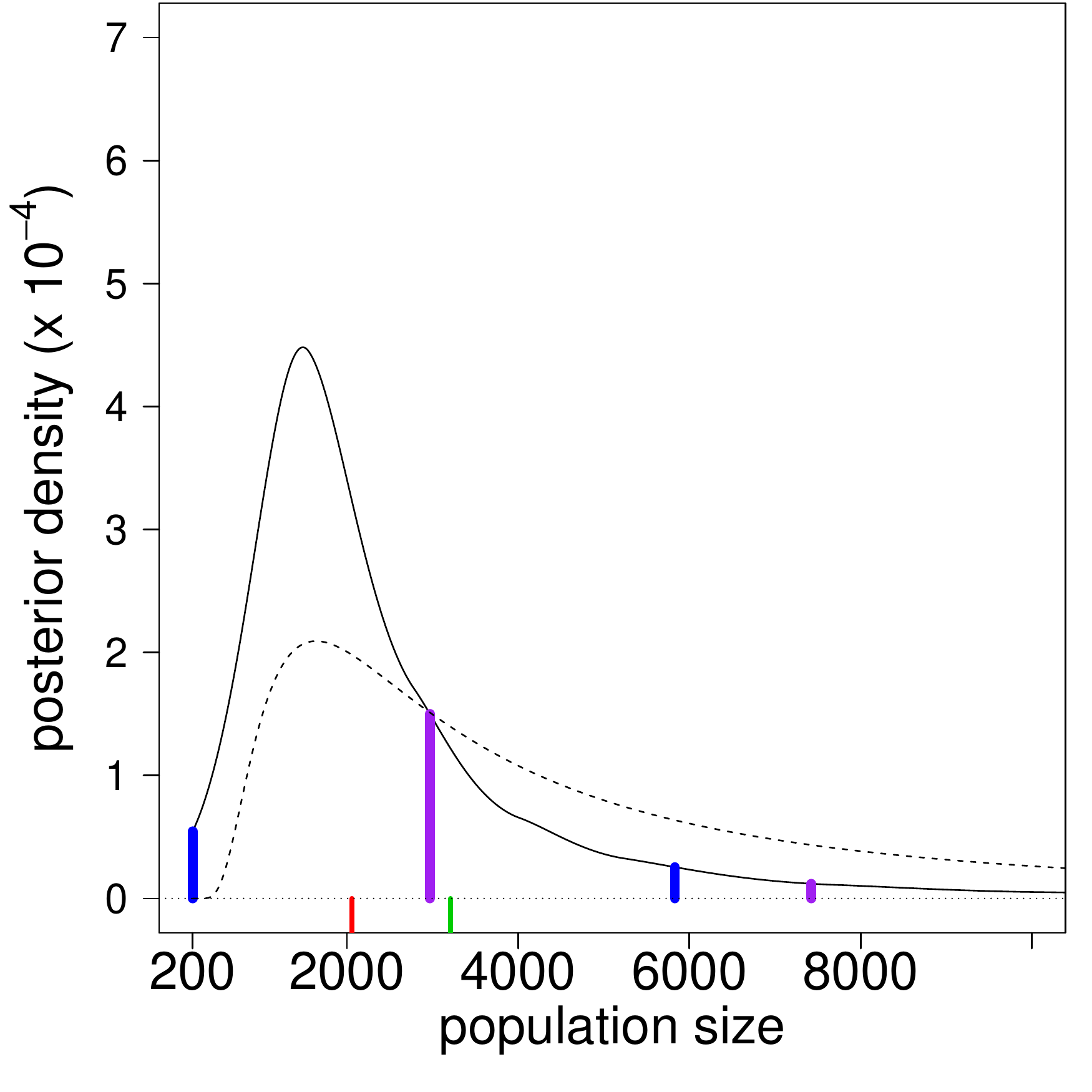}
} \hspace{-.3cm}
\subfigure
%, $Z|Y \sim \delta$}] % caption for subfigure a
{
    \label{smUNAIDSN}
    \includegraphics[width=2.1in]{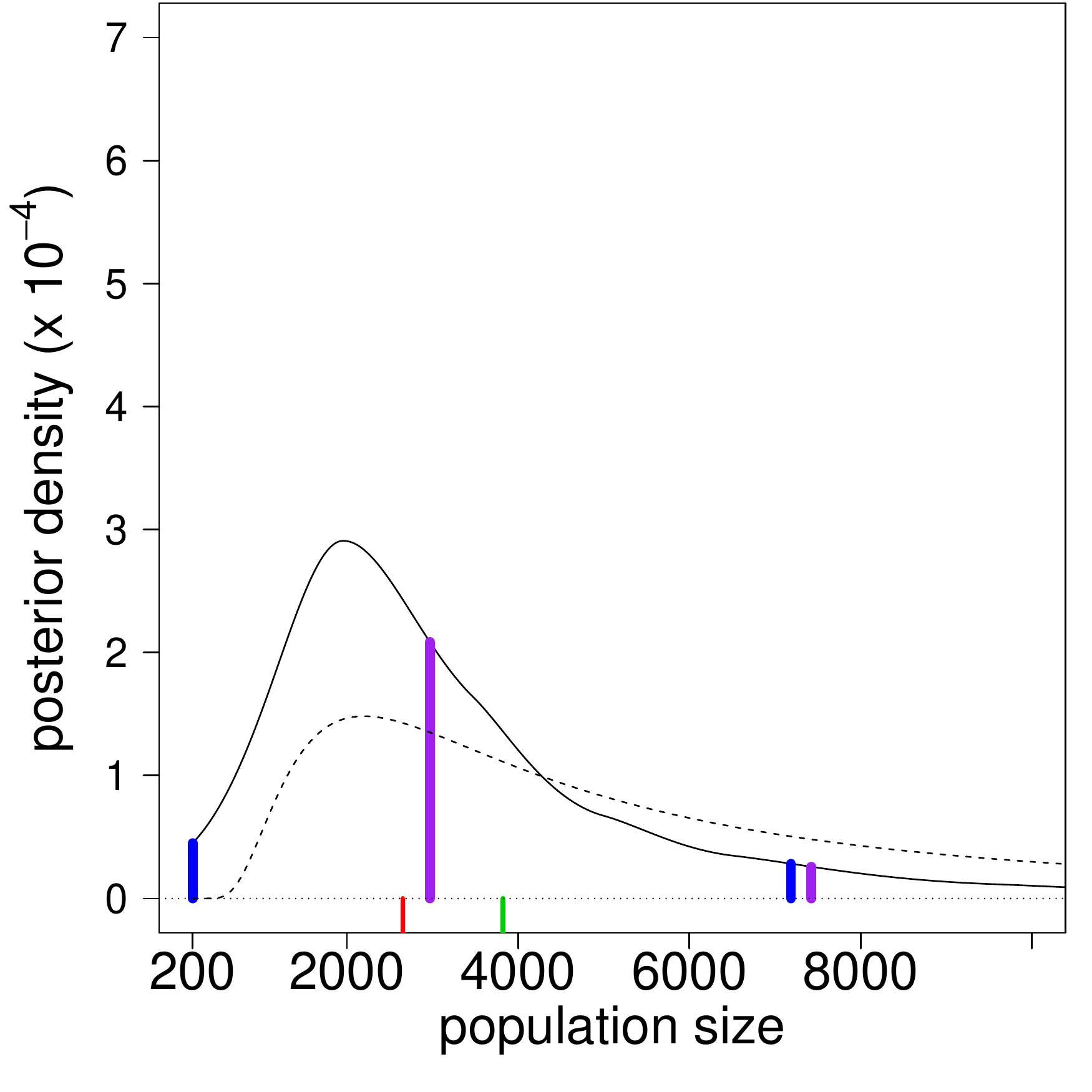}
   } 
%\end{center}
\caption{Posterior distribution for the number of MSM in
San Miguel based on three prior distributions for the population size:  flat, matching the midpoint UNAIDS estimate, and interval-matching the UNAIDS estimate.
The legend is the same as Figure 5.}
\end{figure}

Figure 6 shows the same three prior specifications as the previous case.  Figure \ref{smflatN} plots the posterior distribution and the prior when the prior is constant over the range of population sizes where the likelihood is non-negligible.
The peakedness of the posterior shape again indicates that there is information in the data about the size, but it is diffuse and has a long upper tail compared to that for the FSW. The UNAIDS guidelines (purple lines) fall in the mid to upper part of the distribution, and are well within the 95\% HPD interval (blue lines).

Figure \ref{smmidpointN} plots the  posterior distribution based on the prior with mean the mid-point of the UNAIDS suggested value
$3.5\%\times{148489}=5197$.
The majority of the posterior is below the UNAIDS estimates as the prior 
pulls in the larger values while the 95\% interval mostly covers them.

Figure \ref{smUNAIDSN} plots the posterior distribution based 
on the prior fixing the mean at the midpoint of the range
($3.5\%$) and the lower point ($2\%$) at the lower quartile.
This prior contains the most information from the UNAIDS work group and hence is perhaps the best choice.
The resulting posterior distribution has a mode at about 2100 MSM, and a 95\% HPD interval from
200 to 7048 MSM. Thus this method yields an estimate of the number of MSM in San Miguel with a wide 
 interval.

Another reference point for population size estimates in El Salvadorian cities can be extrapolated from the results of
another study in San Salvador.  In this study, capture-recapture was used to estimate the number
of MSM and FSW in San Salvador in 2008 \citep{Pazbailey2011}.
This approach required the distribution of tokens (e.g., key chains) throughout
the population followed by a recapture step with a follow-up survey.  
\citet{Pazbailey2011} estimate that the size of the FSW population in 
San Salvador is almost double the UNAIDS figures (1.4\%). 
This population proportion in Sonsonate would translate to 2079 FSW, 
close to
the posterior mean of the proposed method for the flat prior, but in the upper tail of the
posteriors based on priors based on the UNAIDS guidelines. 
\citet{Pazbailey2011}  estimate that the size of the MSM population in 
San Salvador is close to the UNAIDS figures (3.4\%). 
This is somewhat high but comparable to the MSM results in Figure 6.
Thus the results here are largely comparable to those of 
\citet{Pazbailey2011} when they differ from the UNAIDS guidelines.

\section{Discussion}

The primary contribution of this paper is a method to estimate population size from RDS data alone.  All existing methods require at least two data sources, and strong assumptions about their dependence structure.  Intuitively, when unit sizes are associated with sampling probability, a systematic decline in observed unit sizes over time is indicative of the depletion of the available population.  As described in this paper, a successive sampling approximation to the RDS process leverages this change in observed sizes to estimate the size of the hidden population.  These data were previously unexploited in the estimation of the size of hard-to-reach human populations.  Because RDS is designed for inference in hard-to-reach populations, such data often exist in precisely the populations where population size is unknown but of great interest.  Thus this method provides additional important information,  that is, an estimate of population size, at no extra cost. Furthermore, the Bayesian framework of this work allows for easy incorporation of informative prior information or data from other sources.

The main limitation of the proposed method is the small amount of information on population size available in many RDS samples, with less information available in smaller sample fractions.  In cases with little information, this method results in very large interval estimates in order to obtain reasonable frequentist coverage properties.  
However, existing methods are also subject to great uncertainty, even though they typically
require additional data collection. Often these methods lead to apparently
conflicting results because they poorly estimate measures of uncertainty
\citep{Salganik15112011}. The advantage of the proposed method is that it uses existing data
and accurately assesses the degree of uncertainty in $N$ over a wide range of
practical situations.  Results from the two cases in El Salvador provide 
estimates compatible with UNAIDS guideline estimates as well as capture/recapture estimates of population sizes.

This method is also useful for estimators of population characteristics that %take into account finite population effects and hence 
 require an estimate of the population size.  The simulation study %described here 
 demonstrates that using population size estimates from the proposed method in the SS estimator \citep{gilesppsjasa09} works well and is particularly helpful in conditions of strong differential activity and larger sample fractions.

The framework developed in this paper is designed to be a foundation
upon which other approaches to population size estimation can build. In
particular it is designed to facilitate combination with data from
multiple methods (e.g., direct surveys, capture-recapture, network scale-up and
multiplier). The posterior from the approach in this paper can be used as a
prior for combination with information from these alternative methods.
Thus this methodology will lead to coherent inference that can be incrementally
improved in a constructive way.

While the methods in this paper have been applied to data collected via RDS, we note that the approach is general and
applies to data collected via successive sampling. Hence the method
has broad applicability \citep{andkau86,nairwang89,bnw92}.

The R package implementing the methods developed in this
paper \blind{\citep{size}}{\citep{sizeblind}}
will be available
on CRAN \citep{R}.

\section{Supplementary Materials}

\begin{description}
   \item[Inferential and Computation:]This supplement presents specifics of the estimation algorithms, variations of models for the unit size distribution, a proof of the non-amenability of RDS, and the recruitment graph for the study of female sex workers in El Salvador.
\end{description}

\blind{{\relax}{
\section*{Acknowledgments}
\rm The project described was supported by grant number 1R21HD063000 from NICHD and
grant number MMS-0851555 from NSF, and grant number N00014-08-1-1015 from ONR. Its contents are solely the responsibility
of the authors and do not necessarily represent the official views of the
Demographic \& Behavioral Sciences (DBS) Branch,
the National Science Foundation, or the Office of Navel Research.
The authors would like to thank the members of the Hard-to-Reach Population Research Group (\url{hpmrg.org}), especially Lisa G. Johnston, for their helpful input.
We would like to express our gratitude to the Ministry of Health of El Salvador for
allowing us to use the results of the Encuesta Centroamericana de VIH y
Comportamientos en Poblaciones Vulnerables, ECVC-EL Salvador. We would especially
like to thank the study principal investigators Dr. Ana Isabel Nieto and 
Gabriela Paz-Bailey, and the study coordinator Maria Elena Guardado for their assistance
interpreting the results of this analysis. Funding for the ECVC-El Salvador study was
provided by the United States Centers for Disease Control and Prevention, United
States Agency for International Development, the Ministry of Health of El Salvador,
and the World Bank.
}}

\addcontentsline{toc}{section}{References}
\medskip
%\bibliographystyle{JASA_manu}
%\blind{{\bibliographystyle{JASA_manu}}
%       {\bibliographystyle{plainnat}}}
\bibliography{biomsize0}
%\begingroup
%\setstretch{0.9}
%\setlength\bibitemsep{5pt}
%\printbibliography
%\endgroup

\end{document}

\maketitle

\setcounter{section}{0}
\renewcommand{\thesection}{S.\arabic{section}}
\setcounter{equation}{0}
\renewcommand{\theequation}{S.\arabic{equation}}
\setcounter{figure}{0}
\renewcommand{\thefigure}{S.\arabic{figure}}

\section{Algorithmic Details} \label{smhsim}

This section derives the algorithm used to compute the joint posterior 
for the population size, the super-population parameter $\eta$ and the unobserved population sizes
$U_{unobs}$. It is based on that developed by \citet{west96}.

First consider the case in Section 2.3 where the population size $N$ is assumed known.
The augmented posterior:
\begin{equation}
p(\eta, U_{unobs} = u_{unobs}, \Psi | D)
\label{saugpost}
\end{equation}
can be computed via a three component
Gibbs sampler.  Explicitly, the steps are:
\begin{enumerate}
\item Initialize $u_{unobs}$ at a set of unit sizes. %(e.g, by re-sampling $u_{obs})$.
\item Sample $\eta$ from
\begin{equation}
p(\eta \vert U_{unobs} = u_{unobs}, \Psi, D)
=
\pi(\eta){\cdot}
\prod_{j=1}^{n}{f(u_{g_j} | \eta)}\cdot
\prod_{j \in \bsg}{f(u_j | \eta)}\label{sampeta}
\end{equation}
\item Sample $\Psi$ from
$p( \psi_{j}=\psi | \eta, U_{unobs} = u_{unobs}, D)$ in equation (7).%(\ref{fcpsi}).
\item Sample $U_{unobs}$ from
$p(U_{unobs} = u_{unobs} \vert \Psi, \eta, D)$
in equation (8).%(\ref{fcdunobs}).
\item Repeat steps (2) through (4) until convergence.
\end{enumerate}

The MCMC should be run until burn-in and then sampled after an interval of
iterations to produce a large sample from the augmented posterior
(\ref{saugpost}). Standard MCMC diagnostics can be applied to assess convergence \citep{gilks.et.al:bk:1996}. In our experience, the procedure is well behaved and converges quickly.
The augmented posterior can be marginalized to produce samples from
$p(\eta \vert D)$ and 
from the posterior predictive distribution for the unobserved unit sizes:
$p(U_{unobs} = u_{unobs} \vert D)$.

These in turn enable inference for such quantities as the mean unit size, the
unit size distribution, etc.

We make the following step-by-step remarks on the algorithm:% in Section \ref{sec:bayesdegree}:
\begin{enumerate}
\item $u_{unobs} \sim f(\cdot | \eta_0),$ where $\eta_0$ is a specified
starting value.
\item This is done with a Metropolis-Hastings algorithm with Gaussian
proposal for the mean size and a inverse $\chi$ for the standard deviation
of the size. 
\item These are independent standard Exponential draws.
\item This is done with a rejection algorithm \citep{west96}. For each element
of $u_{unobs}:$
\begin{enumerate}
\item Draw $d \sim f(\cdot | \eta)$ and independently, $u \sim U(0,1)$. 
\item If $\log(u) > -d$ then reject $d$ and return to (a); otherwise save
$d$ as the element of $u_{unobs}$ and return to (a) for the next element.
\end{enumerate}
\end{enumerate}

In the situation in Section 2.4 where $N$ is not known, we need to extend the above algorithm
to sample from the joint posterior
$p(N, \eta, U_{unobs}, \Psi | D)$.

The algorithm is:
\begin{enumerate}
\item Initialize $N$ at a point estimate and $u_{unobs}$ at a set of unit sizes. %(e.g, by re-sampling $u_{obs})$.
\item Sample $\eta$ from
\begin{equation}
p(\eta \vert U_{unobs} = u_{unobs}, \Psi, D, N)
=
\pi(\eta){\cdot}
\prod_{j=1}^{n}{f(u_{g_j} | \eta)}\cdot
\prod_{j \in \bsg}{f(u_j | \eta)}
\end{equation}
\item Sample $\Psi$ from
$p( \psi_{j}=\psi | \eta, U_{unobs} = u_{unobs}, D, N)$ in equation (7). %(\ref{fcpsi}).
\item Sample $N$ from
$p(N \vert \eta, \Psi, D)$
in equation (13). %(\ref{fcm}).
\item Sample $U_{unobs}$ from
$p(U_{unobs} = u_{unobs} \vert \Psi, \eta, D, N)$
in equation (8).%(\ref{fcdunobs}).
\item Repeat steps (2) through (5) until convergence.
\end{enumerate}

As before, this expanded
MCMC can be run to produce a large sample from the augmented posterior:
\begin{equation}
p(N, \eta, U_{unobs} = u_{unobs}, \Psi | D)
\label{esaugpostN}
\end{equation}
This can then be marginalized to produce samples from
$p(N \vert D),$ 
$p(\eta \vert D)$ 
and the posterior predictive distribution of the unobserved unit sizes, 
$
p(U_{unobs} = u_{unobs} \vert D)
$. Hence it produces posterior predictive distributions of the full population
of unit sizes $(u_i, i=1,\ldots, N)$.

The step-by-step remarks are the same as before. For 
step 4, the complete PMF is discrete and is computed directly 
from (13) %(\ref{fcm}) 
 and hence sampled from directly.

Both these algorithms have been implemented at the {\tt C} level in the 
{\tt R} package {\tt size} \citep{size}.
They are accessible via a user-friendly front-end.
It includes a range of unit size distribution models, each of which 
is parametrized in terms of the mean and
standard deviation of the distribution.

\section{Models for the Unit Size Distribution}\label{ssdm}

The methods in this paper require a parametric model for the super-population
draws of the unit sizes. Although the overall mathematical framework
is general, the case where the unit sizes
are degrees is considered here. 

There is a substantial literature on models for
the degree distributions of social networks 
\citep{joneshandcock2003a,joneshandcock2003b,handcockjones2004,handcockjones2006}. 
The naive models for these distributions include the Poisson and the 
Negative binomial (to allow Gamma over-dispersion relative to the Poisson).
However, degree distribution tend to be long-tailed, suggesting alternatives such as the
Yule and the Waring distributions \citep{joneshandcock2003a}, which allow power-law over-dispersion. Another in this class is the 
Poisson-log-normal, allowing log-normal over-dispersion \citep{perline05}.
This allows longer-tails than the Negative Binomial but less than the power-law models. These models are unable to represent under-dispersion of the degree counts. The Conway-Maxwell-Poisson distribution allows both under-dispersion and
over dispersion with a single additional parameter over a Poisson
\citep{shmuelietal2005}.
It is possible to augment these distributions by directly
parameterizing the lower-tail or upper-tail
(e.g., $f(1|\eta), f(2|\eta), \sum_{k=20}^{\infty}{f(k|\eta)}$). 
This can improve the fit.
Each of these options was considered for this paper and are available in the 
R package {\tt degreenet} on CRAN \citep{degreenet, R}.
% or the R package {\tt size}.
In many applications it is helpful to focus on two-parameter models which enable separate
specification of the location and spread of the degree distribution (with the shape a feature of the
model).

\section{Prior for the population size}

This issue is considered in Section 2.6 of the paper.
In addition to the improper uniform prior, there are natural
classes of parametric models for counts considered for the unit sizes above.
(e.g., Negative Binomial, Poisson-log-normal, Conway-Maxwell-Poisson).

Note that the data effectively truncates the prior below the sample size $n$.

The most common parametric models for $N$ (e.g., Negative Binomial) typically 
have too thin tails for large $N$ and require accurate specification of the prior mean and standard deviation.

The class of priors proposed in Section 2.6 
specifies prior knowledge about
the sample proportion (i.e. $n/N$).
This is based on the idea
that the sample size may not be chosen 
separately from the population size but is
often chosen in relation to it.  So a simple prior is a uniform prior on the
sampling proportion (i.e. the proportion of the population in the sample).
This translates to a closed form for the prior on $N$ which has 
infinite mean (and higher moments) and allows for very large population sizes.
To allow a more flexible prior, we specify a Beta$(\alpha,\beta)$ distribution 
on the sample proportion. This is translated to a prior on the discrete
support of $n/N$ by assigning the probability mass to the closest discrete
value. 

A uniform distribution on the sample proportion corresponds to a median of
twice the sample size.
While the mapping from the mean to $\beta$
does not have a closed from (for general $\beta$) it can be
easily computed numerically.
We have found the sub-class with $\alpha=1$ to be the most useful. For these 
the mode of the prior is at $0.5{n}(\beta+1)$
and the median is given by
$n / (1-(1/2)^{1/\beta})$.

\section{RDS is not amenable to the model} \label{amenable}

The RDS design is not amenable to the modeling framework used in this paper, and here we explain why.
In Section 2.2, modeling the sampling design along with the resulting
data structure adds a great deal of complexity (see, e.g., (4)).  It is natural to ask when we might consider the simpler face-value likelihood,
\begin{equation}
L[\eta \vert U_{obs} = u_{obs} ]\propto 
\prod_{j=1}^{n}{f(u_j | \eta)}\cdot\sum\limits_{u_{unobs} \in {\cal U}(u_{obs})}{\prod_{j=n+1}^{N}{f(u_{j} | \eta)}},
\label{facevalue}
\end{equation}
which ignores the sampling design.

The sampling design is amenable, in the sense of \cite{hangile10aoas}, if:
\begin{eqnarray*}
P(G=g_{obs} \vert U_{obs} =u_{obs}, U_{unobs} =u_{unobs}, \eta) = 
P(G=g_{obs} \vert U_{obs} =u_{obs}).
\end{eqnarray*}

If the sampling design is adaptive:
\begin{equation*}
L[\eta \vert U_{obs} =u_{obs}, G=g_{obs}]
\propto P(G=g_{obs} \vert U_{obs}
=u_{obs})\cdot
L[\eta \vert U_{obs} =u_{obs} ]
\end{equation*}

Thus if the sampling design is adaptive then the sampling design is
\textit{ignorable} in the sense that the resulting likelihoods
are proportional \citep{litrub2002}.
When this condition is satisfied
likelihood-based inference for $\eta,$ as proposed here, is
unaffected by the (possibly unknown) sampling design.

However, based on equation (2), %(\ref{sstransition}), 
 the sequential
sampling design is {\sl clearly not adaptive}. Likelihood inference should
be based on the full likelihood given by equation (1). %(\ref{fullikelihood}).

\section{Recruitment Graph of the Female Sex Workers in Sonsonate}\label{fswrecruitment}

\begin{figure}[h]
\centering
%\begin{center}
    \includegraphics[width=6.5in,height=2.0in]{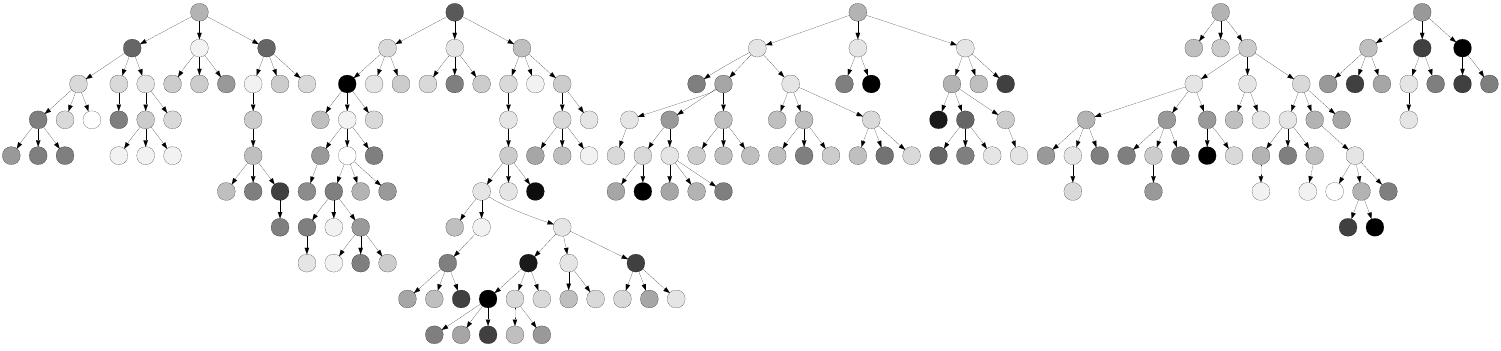}
%\end{center}
\caption{Graphical representation of the recruitment tree for the sampling of female sex workers in
Sonsonate, El Salvador in 2010. The nodes are the respondents and the wave number increases as you go down the page.
The node gray scale is proportional to the network size reported by the worker,
with white being degree one and black the maximum degree.}
\label{sonrecruitment}
\end{figure}

Figure \ref{sonrecruitment} 
is a graph of the 
recruitment tree for RDS of female sex workers in 
the department of Sonsonate, El Salvador. The data is considered in depth in
Section 4.

\bibliography{biomsize0}